\documentclass[preprint,12pt]{elsarticle}

\usepackage{amssymb,amsmath}
\usepackage{graphicx}
\usepackage{dcolumn}
\usepackage{bm}
\usepackage{color}
\usepackage{float}
\usepackage{hyperref}
\usepackage[version=4]{mhchem}
\usepackage{relsize}
\usepackage{comment}
\usepackage{media9}
\usepackage{dblfloatfix}    
\usepackage{textcomp}
\usepackage{xurl}

\newcommand{\icol}[1]{
  \left[\begin{smallmatrix}#1\end{smallmatrix}\right]}

\journal{XXX}

\begin{document}

\begin{frontmatter}



\title{The Physics of Preference: Unravelling Imprecision of Human Preferences through Magnetisation Dynamics}


\author[label1]{Ivan S.~Maksymov}

\affiliation[label1]{organization={Artificial Intelligence and Cyber Futures Institute, Charles Sturt University},
            city={Bathurst},
            postcode={2795}, 
            state={NSW},
            country={Australia}}

\author[label1,label2,label3]{Ganna~Pogrebna}

\affiliation[label2]{organization={The Alan Turing Institute, British Library},
            city={London},
            postcode={NW1 2DB}, 
            country={United Kingdom}}
            
\affiliation[label3]{organization={The University of Sydney Business School},
            city={Darlington},
            postcode={2006}, 
            state={NSW},
            country={Australia}}

\begin{abstract}
Paradoxical decision-making behaviours such as preference reversal often arise from imprecise or noisy human preferences. Harnessing the physical principle of magnetisation reversal in ferromagnetic nanostructures, we developed a model that closely reflects human decision-making dynamics. Tested against a spectrum of psychological data, our model adeptly captures the complexities inherent in individual choices. This blend of physics and psychology paves the way for fresh perspectives on understanding the imprecision of human decision-making processes, extending the reach of the current classical and quantum physical models of human behaviour and decision-making.
\end{abstract}




\end{frontmatter}


\section{Introduction}
Theoretical and experimental research in economics and psychology posits that human preferences are inherently noisy and imprecise \cite{Lin71} (see \cite{Bha17, loomes2017preference, blavatskyy2021probabilistic, busemeyer2019primer, sharma2023scarcity} for definitions of noise and imprecision as well as detailed reviews of the relevant literature). For example, such phenomena as \textit{Allais Paradox}, \textit{Ellberg Paradox}, \textit{Endowment Effect} and \textit{Preference Reversal} (PR) demonstrate that human decisions deviate from predictions of traditional rational models \cite{moffatt2009experimental}. These deviations follow certain patterns \cite{loomes2017preference}. While the classical theory suggests that individuals consistently maximise their utility based on stable preferences, experiments present robust evidence that human choices under risk and uncertainty are often inconsistent or even contradictory \cite{chater2018mind, blavatskyy2010models, bardsley2009experimental}.

PR was first observed in the gambling studies involving \cite{Lic71}: (i) a choice task, where people were asked to decide between a relatively safe lottery with a low payoff (Probability- or P-bet) and a relatively risky lottery with a high payoff (Dollar- or \$-bet) as well as (ii) a valuation task, where the same people were asked to indicate lottery prices if they were to sell P-bet and \$-bet. The observed people’s preferences were often paradoxical: while opting for the P-bet in the binary choice, they were putting a higher price on the \$-bet, thereby indicating a preference for P-bet in a simple two-option choice, but preference for \$-bet when stating their certainty equivalents for both options.

Alongside other behavioural irregularities (see, e.g.,~\cite{bardsley2009experimental, blavatskyy2023common}), PR has reshaped contemporary discourse in both economics and psychology. Traditional deterministic models such as Expected Utility Theory (EUT), which held that individual choices were reflective of consistent and stable preferences, have been challenged by empirical findings \cite{Gre79, loomes2017preference}. Furthermore, recent research demonstrated that preference reversals constitute a challenge for many stochastic decision theories (e.g.,~when decision theories such as Expected Utility Theory or Cumulative Prospect Theory are embedded into models of stochastic choice such as Fechner model) \cite{butler2007imprecision, loomes2017preference}. The latter approach has demonstrated that individual choices under risk are probabilistic rather than deterministic, i.e.~the same person may make different choices when presented with the same problem multiple times. Experiments show that people's decisions to accept or reject sure payoff amounts in exchange for a risky lottery varies from consistently selecting bet over cash (when the sure payoff amounts offered against the lottery are not sufficiently attractive) to consistently selecting cash over bet (when the sure payoff amounts increase) with some \textit{imprecision interval} in-between, when individuals make probabilistic choices between cash and the bet \cite{mosteller1951experimental, butler2007imprecision,loomes2014testing, loomes2017preference}. The existence of such paradoxes implied that human decision-making was more multifaceted than previously assumed.

Understanding and accurately predicting these anomalies is not merely an academic exercise. As policy-makers strive to devise strategies and interventions that depend on opinion polls, the need for models predicting human behaviour is pressing. Yet, capturing the essence of phenomena like PR within traditional deterministic and stochastic models remains a challenge. For example, many deterministic and stochastic decision theories honour the transitivity axiom, which posits that if an individual prefers $A$ to $B$ and $B$ to $C$, then they should prefer $A$ to $C$. However, PR defies the transitivity, suggesting that human preferences might be imprecise \cite{butler2007imprecision, butler2018predictably}.

\begin{figure}
 \includegraphics[width=0.79\textwidth]{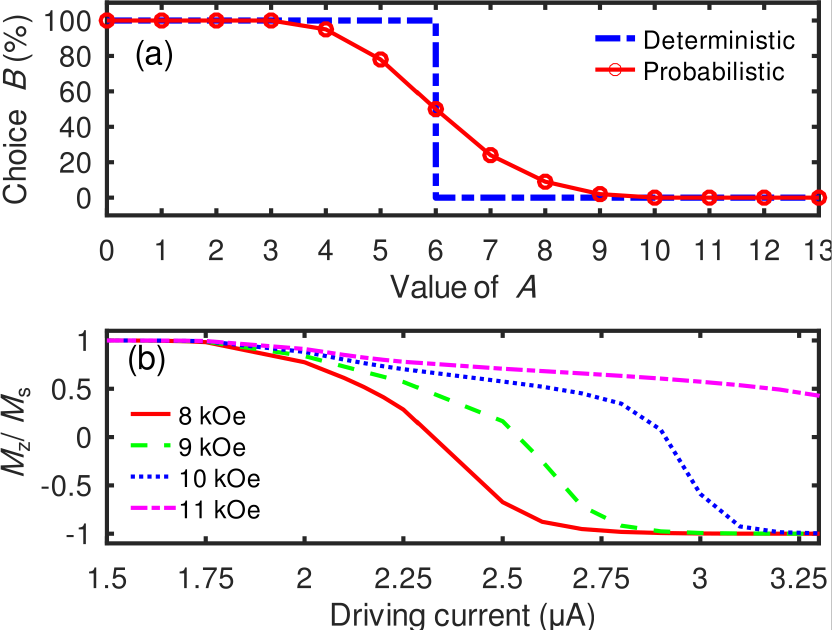}
 \caption{(a)~Idealised deterministic (the dashed line) and probabilistic (the line with circular markers) PR curves. Examples of real-life data that resemble the shape of the idealised curve are given in Fig.~\ref{Fig3}. (b)~Representative simulated magnetisation reversal behaviour for different values of the static magnetic field $H_0$. Additional physical details are presented in the main text and the caption to Fig.~\ref{Fig4}. The shape of the probabilistic curve in panel~(a) closely resembles the magnetisation curves in panel~(b). Either a single magnetisation reversal curve or a judicious combination of several magnetisation reversal curves, as demonstrated in the main text, can approximate both experimental curves in Fig.~\ref{Fig3} and more complex dataset in Fig.~\ref{Fig5}.\label{Fig1}}
\end{figure}

This instigates debates on the need to (i)~reconsider some axioms of decision theory \cite{butler2018predictably, blavatskyy2021probabilistic}; (ii)~devise theories relaxing transitivity assumption \cite{loomes1982regret}; (iii)~embed decision theories into models of noise-synthesising stochastic versions of known theories \cite{blavatskyy2010models}; and (iv)~develop further models that assume inherently imprecise human preferences \cite{loomes2017preference}.
\begin{figure}
 \centering
 \includegraphics[width=1\textwidth]{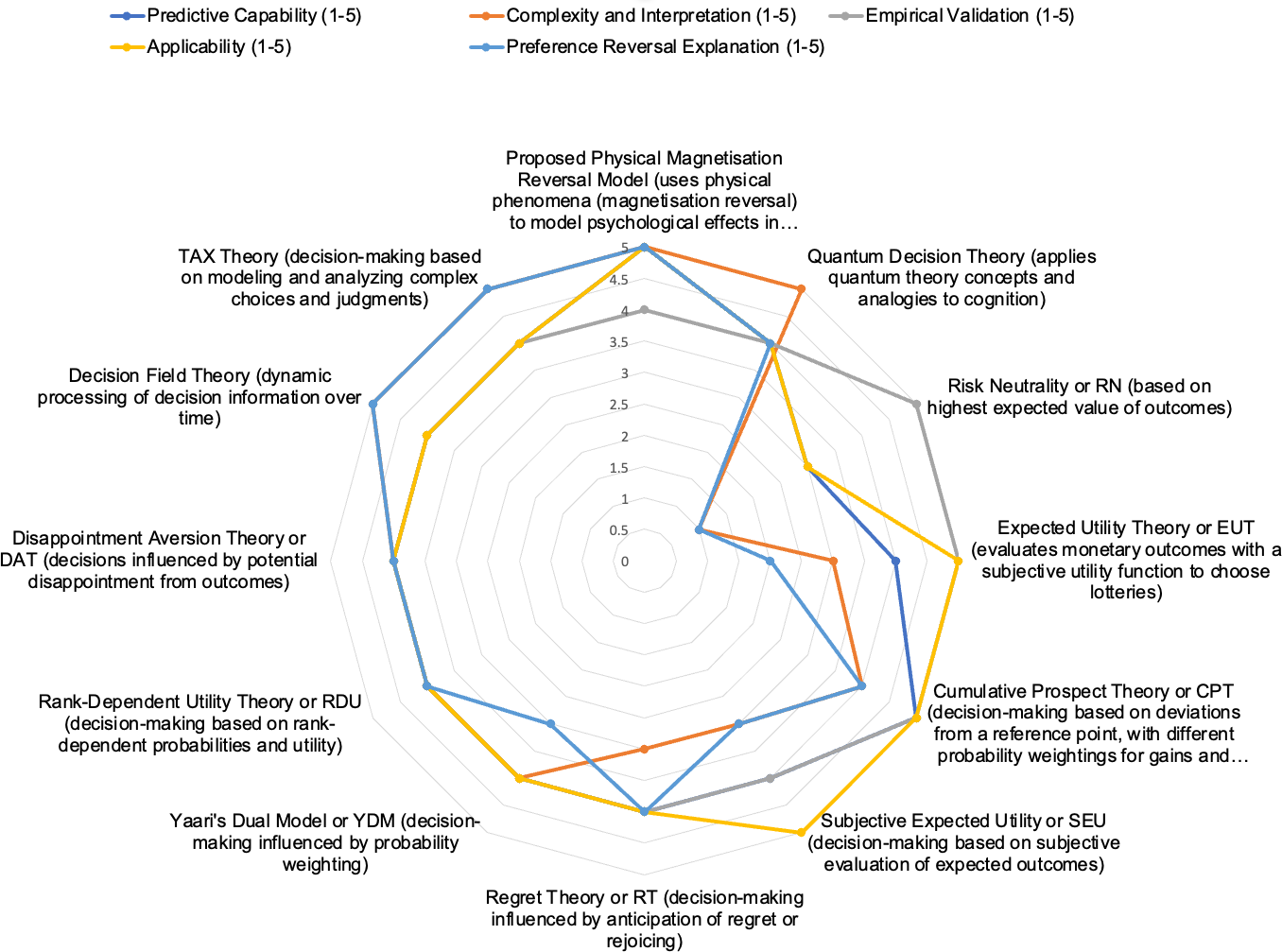}
 \caption{Location of the proposed physical Magnetisation Reversal model on a diagram that evaluates the existing decision theory models using a 1--5 scale applied to such model characteristics as Predictive Capability, Applicability, Complexity and Interpretation, Preference Reversal Explanation and Empirical Validation. To obtain the relative scoring, we first compiled a detailed table of decision theories based on the comprehensive literature review and the content of the relevant papers. We then assigned a score based on table entries and plotted the resulting diagram based on the obtained scores. This diagram does not reflect the fact that the Magnetisation Reversal model is physically phenomenological, was developed independently of any other decision theory model and, {\it de facto}, is unique as further demonstrated in Fig.~\ref{Fig2bis}. The software and raw data used to prepare Fig.~\ref{Fig2} can be accessed following the instructions given in the Data Availability section.\label{Fig2}}
\end{figure}

While these debates are ongoing, it has been noticed that physical processes can adequately describe many psychological effects and decision-making phenomena (e.g.,~quantum logic has been previously applied by social psychologists to model decision making under risk and uncertainty using models of quantum cognition \cite{yearsley2016quantum, busemeyer2019primer}). Furthermore, there exist classical physical models of confirmation bias \cite{Gro17}, public opinion formation \cite{Gal05, Cas09, Red19}, preference \cite{Cap21, Kva21} and backfire effect \cite{Hoh23}. Thermodynamic models of decision-making where information processing is represented as changes in states of a thermodynamical system and quantified by differences in free energy have also been proposed \cite{Ort13, Pak13, Eva21, Ann22}.

Quantum-mechanical models of decision-making \cite{Atm10, Aer14, Bus12, Aln17} that originate from the quantum mind hypothesis \cite{Khr06, mindell2012quantum} have been developed following a suggestion that the human mental states can be represented as a combination of two binary states `0' and `1' \cite{Atm10, Bus12, Pot22}. Mathematically and physically, such a representation corresponds to a superposition of the basis quantum-mechanical states $|0\rangle$ and $|1\rangle$ employed in quantum computers.    

Instead of binary bits, a quantum computer uses a quantum bit (qubit) that corresponds to the states $|0\rangle = \icol{1\\0}$ and $|1\rangle = \icol{0\\1}$. Unlike the `0' and `1' binary states of a classical digital computer, the states of a qubit are a superposition $|\psi\rangle = \alpha |0 \rangle + \beta |1 \rangle$ with $|\alpha|^2 + |\beta|^2 = 1$. As part of a quantum measurement, a qubit interacts in a controlled way with an external system from which the state of the qubit under measurement can be obtained. For instance, using the projective measurement operators $M_0 = |0\rangle\langle0|$ and $M_1 = |1\rangle\langle1|$ \cite{Nie02}, the measurement probabilities for $|\psi\rangle = \alpha |0 \rangle + \beta |1 \rangle$ become $P_{|0\rangle} = |\alpha|^2$ and $P_{|1\rangle} = |\beta|^2$. This measurement outcome means that the qubit will be in one of its basis states.

Quantum computational algorithms are exponentially faster than any possible classical computational algorithm \cite{Nie02}. Therefore, it has been shown that quantum mechanics can explain certain psychological and decision-making processes better than any classical model \cite{Pot09, Bus12, Pot22}. Those theoretical findings have been convincingly supported by a large and growing body of experimental research works that have examined diverse paradoxical phenomena of human behaviour such as Allais Paradox and Ellsberg Paradox \cite{Aer11, Aer12, Blu16, Wei19, Ish20, Ost22, Abe23}.

In this paper, we further extend the ability of physical models of decision-making to capture the effect of {\it preference imprecision} which often results in the preference reversals. In this pursuit, our paper not only makes a completely new connection between physics and decision theory, which has not been explored in the previous literature, but also provides a comprehensive evaluation of a real-life physical model against rich empirical datasets. In particular, apart from several traditionally designed decision-theoretic laboratory experiments involving lotteries, we validate our model using a dataset obtained from the ``Deal or No Deal'' video game based on the famous TV show.

As shown in Fig.~\ref{Fig2}, while our \textit{Magnetisation Reversal} model is capable of explaining phenomena previously discussed in decision-theoretic literature, it makes a novel contribution and is based on completely different core principle contrasting to other theories. Given the novelty of our approach, the whole Fig.~\ref{Fig2} should be viewed mostly as an attempt to find a tentative place for our model in a general taxonomy of the existing decision theory models. For instance, although in Fig.~\ref{Fig2} we place our model near the quantum models of decision-making, strictly speaking our model is classical since the main equation of our model---the Landau-Lifshitz-Gilbert equation---is a classical phenomenological model that describes the quantum-mechanical phenomenon of magnetisation \cite{Lak11}.

By definition, a phenomenological model seeks to represent the essential behaviour observed in the experiment not necessarily attempting to elucidate the origin of the effect using the existing theories \cite{Mcm68}. Hence, our phenomenological model is concerned with reproducing certain decision-making data as they appear in our experience (e.g.,~visually) using the physical effect of magnetisation reversal. Specifically, we choose this effect because the magnetisation reversal curves closely resemble empirical decision-making data discussed in this paper (e.g.,~compare the shapes of the probabilistic curve in Fig.~\ref{Fig1}a and the magnetisation curves in Fig.~\ref{Fig1}b).  

\begin{figure}
 \centering
 \includegraphics[width=1.2\textwidth]{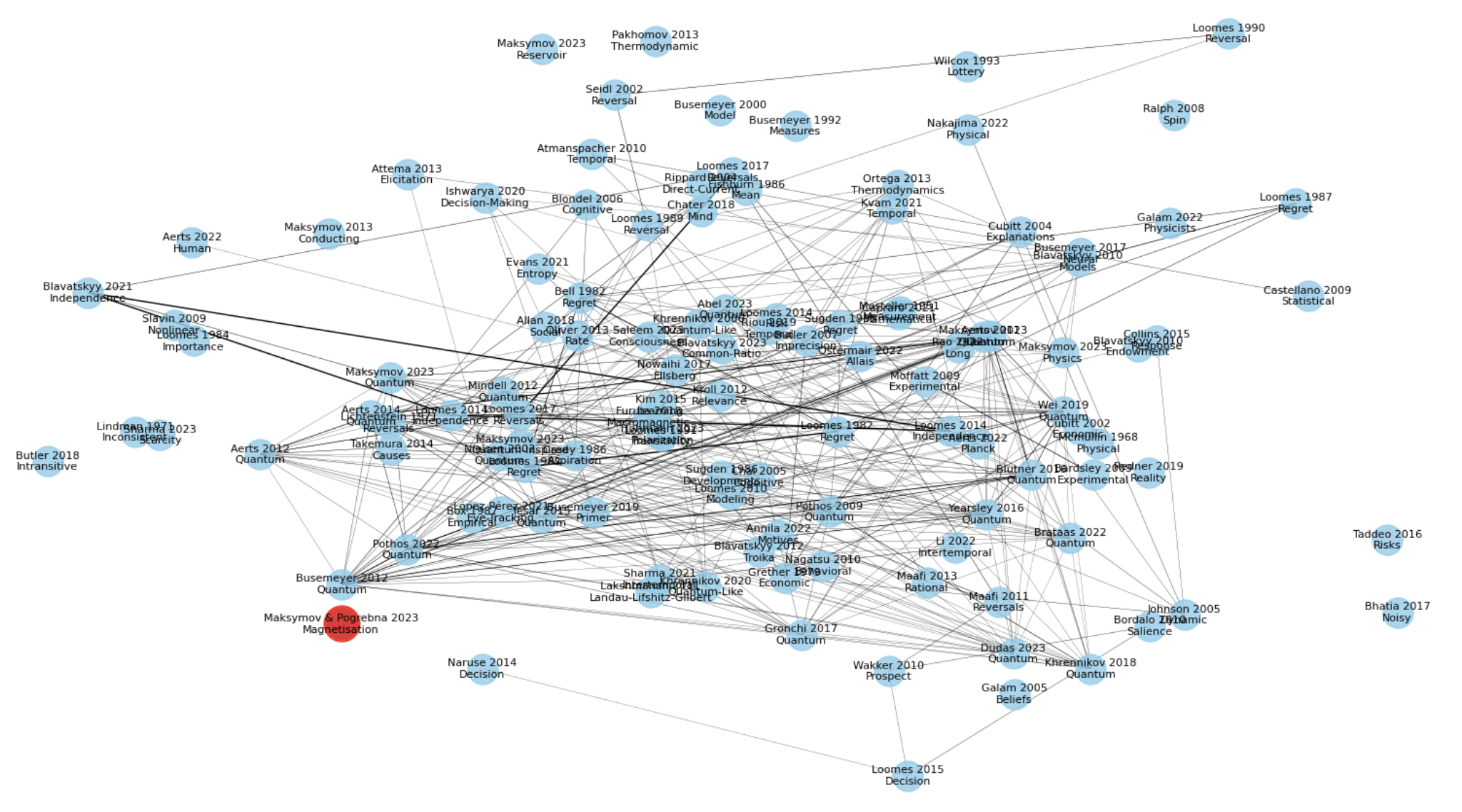}
 \caption{Literature analysis-based taxonomy of decision making models complementing the data presented in Fig.~\ref{Fig2}. The connection chart confirms the core-principle independence of our model (shown in red) of the previous theoretical approaches and establishes our model's distinctiveness compared to other models. The graph demonstrates that while this paper deals with the same behavioural concepts as the previous literature (preference reversals), our approach is different from previously proposed in the literature. The network graph was produced in Python programming language by constructing a similarity index between the papers cited in this manuscript, based on keywords. To ensure robustness, results of our analysis for a subset of papers were cross-checked using the online Connected Papers tool.\label{Fig2bis}}
\end{figure}

Thus, we leave it up to the readers to make either a further connection between our model and the other models or classify our model as a new class of models of decision-making. To help the readers in navigating the literature, in Fig.~\ref{Fig2bis} we visualise the links of our work to the previous relevant publications. We can see that our theoretical and computational approach stands out from the plethora of competing models reported in the literature.

\section{Study~1}
\subsection{Empirical Data and Experimental Procedures}
In several relevant studies recording imprecision intervals, \cite{mosteller1951experimental, butler2007imprecision, loomes2017preference}, an individual is given a number of choices between a lottery $B$ and a series of increasing sure payoff amounts $A_j$,~$j=0,\dots,13$. Since each choice is made multiple times and independently of the previous ones, we model this individual’s preferences as a probability distribution. 

Figure~\ref{Fig1}a shows two theoretically possible distributions. The dashed line denotes the deterministic preferences of an individual who chooses $B$ when $j<6$, never chooses $B$ when $j>6$ and is \textit{exactly indifferent} between $A$ and $B$ when $j=6$. The curve with the circular markers denotes the probabilistic preferences. When the sure thing $ST\leq A_3$, the individual always chooses $B$. When $ST\geq A_{10}$, the individual never chooses $B$. For $A_3<A<A_{10}$ there is an \textit{imprecision interval}.

\begin{figure}
 \includegraphics[width=0.99\textwidth]{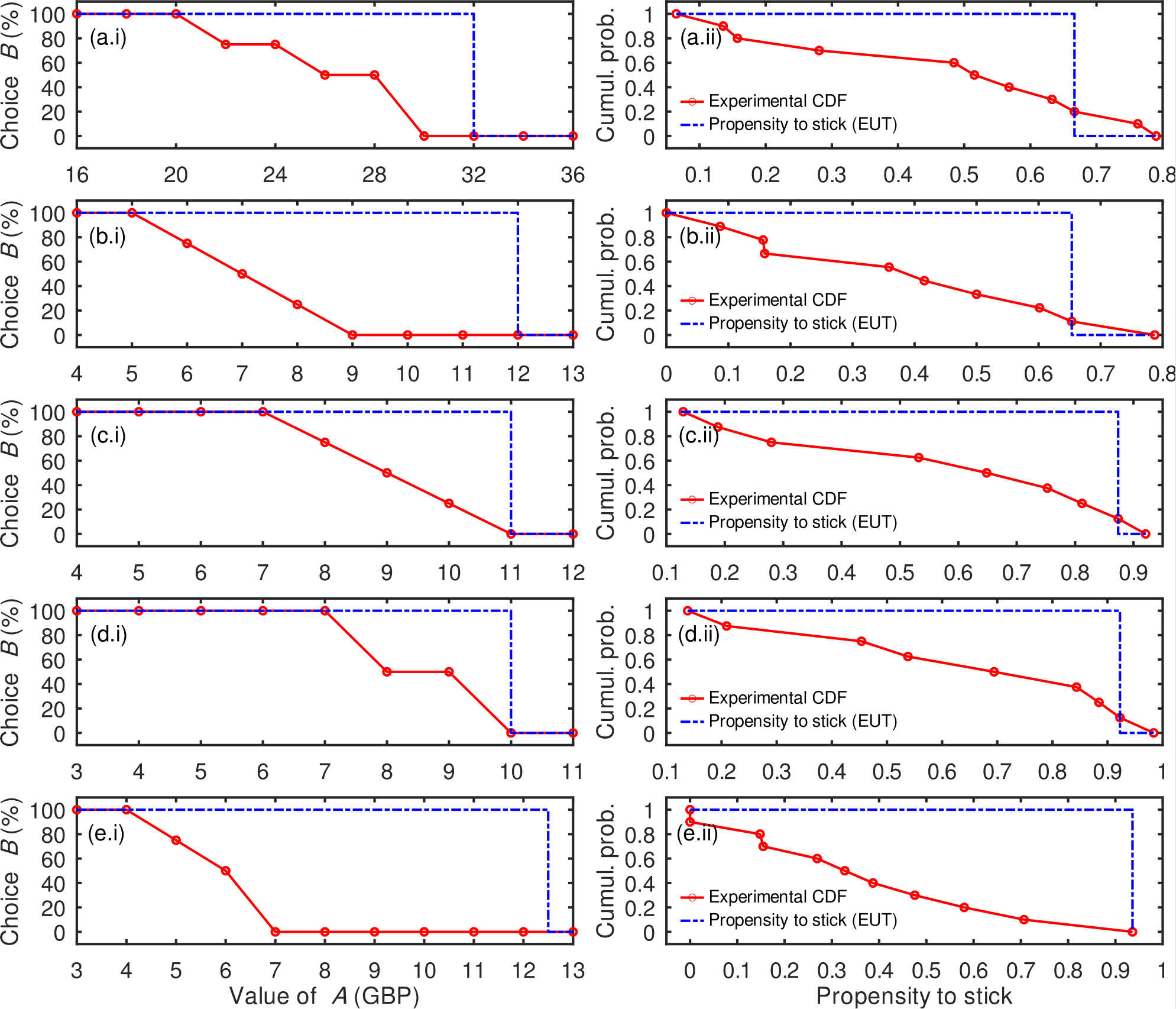}
 \caption{Experimental PR data of the five lotteries (each row of the figure corresponds to an independent dataset). The left column (panels~(a.i)--(e.i)) shows the median patterns, but the right column (panels~(a.ii)--(e.ii)) presents the respective cumulative distributions. Cumulative probability is abbreviated as 'Cumul.~Prob.', CDF stands for cumulative distribution function. The straight dashed lines are the prediction of EUT under the assumption that the participants are risk-neutral. The EUT predictions are provided for illustration only and are not used to validate the magnetisation reversal model. In each row, we first (a.i--e.i) demonstrate the median propensity to stick to a lottery against a series of sure monetary amounts displayed on the horizontal axis (to that end, the number of choices to stick with the lottery was summed across four repetitions for the same subject and median probability to stick with the lottery was calculated across all subjects). On the right hand side (a.ii--e.ii), we demonstrate the corresponding cumulative distribution functions reflective of the mean propensity to stick with the lottery against a series of stakes.\label{Fig3}}
\end{figure}

The probabilistic distribution in Fig.~\ref{Fig1}a is drawn as a sigmoid-like curve since this function robustly fits many intuitions \cite{loomes2017preference} and datasets \cite{mosteller1951experimental, loomes2014testing, loomes2017preference, blavatskyy2010models, butler2018predictably}. To provide a specific example of data resulting in sigmoid-like functions, establishing the robustness of the observed behavioural patterns (sigmoid-shaped curves) in Fig.~\ref{Fig3} we present five examples of experimental curves plotted using the data obtained from the laboratory decision-making experiments with human subjects. We plot the median patterns in the left column of Fig.~\ref{Fig3} and present their respective cumulative distributions in the right column. The Cumulative Distribution Function (CDF) represents the probability that a subject's propensity to stick with a lottery is less than or equal to a certain level as the monetary stakes change as described below.

All experiments were conducted in the University of Warwick using subject pools of undergraduate, postgraduate and executive students over the age of~18 (no minors participated in the experiments). Upon arrival to the laboratory, each participant was provided with the study information sheet and a consent form. Informed consent was obtained from all participants in the study.

In Fig.~\ref{Fig3}a, 81~subjects were making a series of binary choices between a lottery which gave \pounds40 with probability 80\% or 0 with probability 20\% and sure payoff amounts between \pounds16 and \pounds36 with a step of \pounds2. In Fig.~\ref{Fig3}b, 101\, subjects were offered a lottery which yielded \pounds40 with probability 30\% or 0 with probability 70\% versus sure amounts of money between \pounds4 and \pounds12 with a step of \pounds1. In Fig.~\ref{Fig3}c, 101 experimental participants make choices between a lottery providing \pounds15 with probability 70\% and 0 otherwise against sure amounts between \pounds4 and \pounds12 with a step of \pounds1. In Fig.~\ref{Fig3}d, 184 subjects face a choice between a chance of winning \pounds12 with probability 80\% and sure amounts from \pounds3 to \pounds11 with a step of \pounds1. In Fig.~\ref{Fig3}e, 184 people decide between \pounds15 with probability 25\% and sure monetary amounts between \pounds3 to \pounds11 with a step of \pounds1. In all lotteries, each person made each of the binary choices four times. We can see that all five examples demonstrate robust evidence in favour of the existence of an \textit{imprecision interval}, where people make probabilistic choices between a lottery and sure monetary amounts.

Thus, Fig.~\ref{Fig3} plots summary patterns for 81, 101, 101, 184 and 184 experimental participants, respectively. Considering that in the panels (a)--(e) of Fig.~\ref{Fig3} each participant answered at least nine binary choice questions that were repeated four times (hence, we have at least 36 points per participant per case), we additionally plotted the response curves as well as the utility functions for each individual separately, also calculating the number of individuals in each of the cases (a)--(e), who demonstrated patterns consistent with having an imprecision interval according to our definition. Overall, our calculations reveal that in each case the majority of participants exhibited patterns consistent with having an imprecision interval. Importantly, 73\% (59 out of 81), 73\% (74 out of 101), 90\% (91 out of 101), 98\% (181 out of 184) and 93\% (172 out of 184) of participants had patterns consistent with imprecision intervals in the cases labelled as (a),~(b),~(c),~(d) and (e), correspondingly. Raw data relevant to Fig.~\ref{Fig3} can be accessed following the instruction given in the Data Availability section.

\subsection{Magnetisation Reversal Model}
\begin{figure}
 \includegraphics[width=0.7\textwidth]{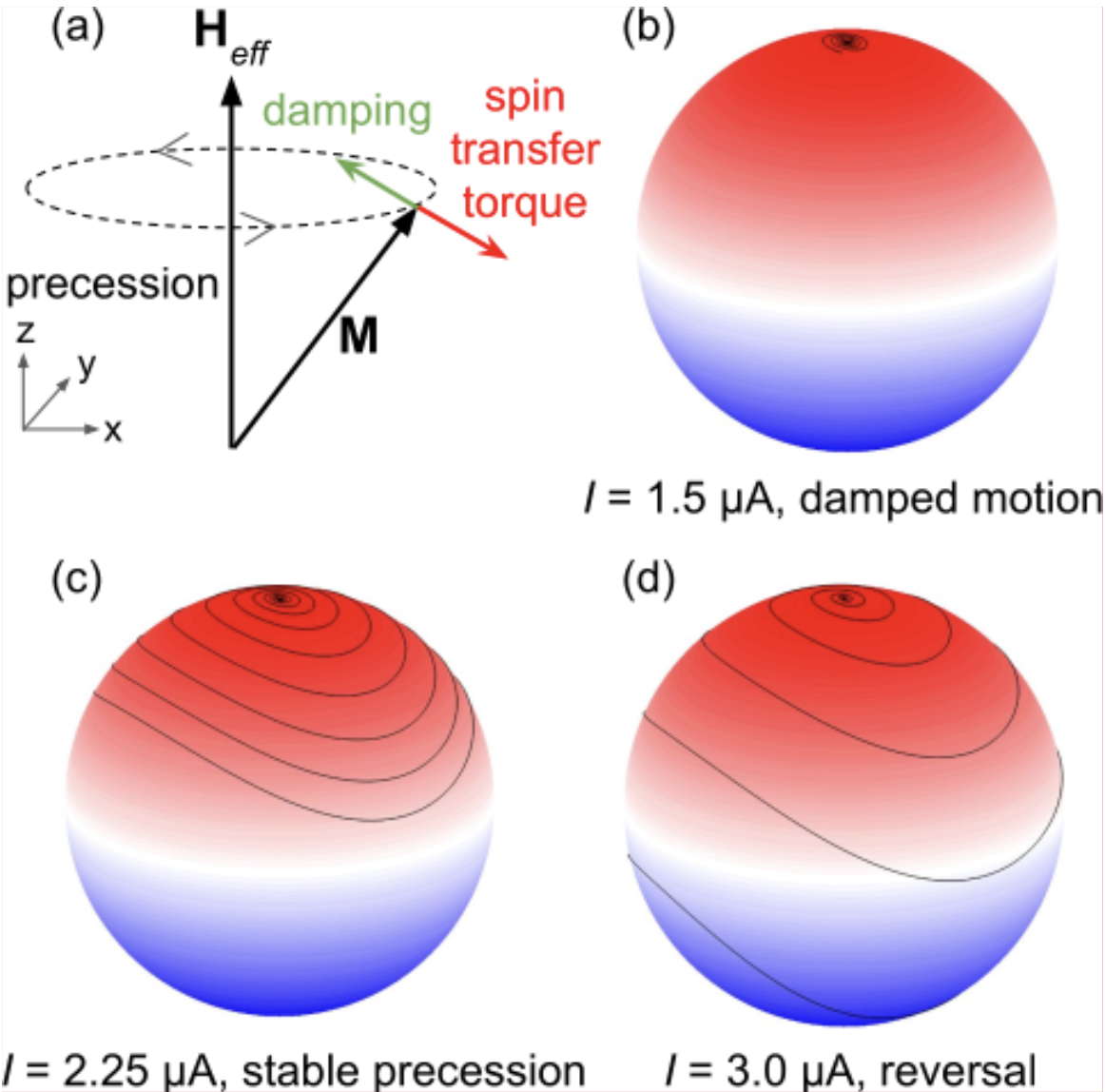}
 \caption{(a)~Sketch of magnetisation vector $M$ precession. Simulated results:~(b)~${\bf M}$ spirals back toward the direction of ${\bf H}_{eff}$ due to the damping. (c)~The spin transfer torque compensates the damping, resulting in a stable precession of ${\bf M}$ around ${\bf H}_{eff}$. (d)~Reversal of the direction of ${\bf M}$. The applied field is $H_0=8$\,kOe. The sphere is used for visualisation only.\label{Fig4}}
\end{figure}

To simulate the experimental decision-making behaviour, we employ magnetisation dynamics in ferromagnetic metal (FM) structures. In general, an electric current is unpolarised since it involves electrons with a random polarisation of spins. However, a current that passes through a thin FM film with a fixed magnetisation direction can become spin-polarised since spins become oriented predominantly in the same direction \cite{Ral08}. This physical effect is exploited in a spin transfer torque nano-oscillator (STNO), where a layer with a fixed direction of magnetisation ${\bf M}$ is separated from a thinner FM layer by a non-magnetic metal layer \cite{Rip04}. When a spin-polarised current is directed from the ``fixed'' magnetisation layer into the ``free'' magnetisation layer, the static equilibrium orientation of magnetisation in the ``free'' layer becomes destabilised. Depending on the strength of the electric current, the destabilisation of magnetisation can lead to either stable precession of magnetisation of the ``free'' layer about the direction of the effective magnetic field or to a reversal of the magnetisation direction (Fig.~\ref{Fig4}).

We model the dynamics of magnetisation using the Landau-Lifshitz-Gilbert equation \cite{Sla09}:
\begin{equation}
  \label{eq:eq1}
  \partial{\bf M}/\partial t = \gamma\left[{\bf H}_{eff}\times\bf M\right] + {\bf T}_G + {\bf T}_{SB}\,, 
\end{equation}
where $\gamma$ is the gyromagnetic ratio. The first term of the right-hand side of Eq.~(\ref{eq:eq1}) governs the precession of $\bf M$ (Fig.~\ref{Fig4}a) about the direction of the effective magnetic field $\textbf{H}_{eff}=H_{0}\textbf{e}_{\rm{z}} + \textbf{H} + \textbf{H}_{ex}$, where $H_{0}$ is the external static magnetic field orientated along the \textit{z}-axis, $\textbf{H}$ is the dynamic field due to currents and magnetic sources such as demagnetising field and eddy current fields and $\textbf{H}_{ex}$ is the exchange field \cite{Mak13}. The dissipative torque is \cite{Sla09}
\begin{equation}
  \label{eq:eq2}
  {\bf T}_G = \alpha_G M_s^{-1}\left[{\bf M}\times{\partial\bf M}/{\partial t}\right]\,, 
\end{equation}
where $M_s$ is the saturation magnetisation of the ``free'' magnetisation layer and $\alpha_G$ is Gilbert damping parameter \cite{Mak13, Sla09}. The Slonczewski-Berger torque is 
\begin{equation}
  \label{eq:eq5}
  {\bf T}_{SB} = \sigma_0IM_s^{-1}\left[{\bf M}\times\left[{\bf M}\times\hat{e}_p\right]\right]\,, 
\end{equation}
where $I$ is the strength and $\hat{e}_p$ is the direction of the spin polarisation of the current. The parameter $\sigma_0$ incorporates fundamental physical constants and the thickness of the ``free'' magnetisation layer \cite{Sla09}.

Equation~(\ref{eq:eq1}) is solved consistently with Maxwell's equations using a finite-difference time-domain (FDTD) method and material parameters presented in \cite{Mak13}. We model a point-contact STNO structure experimentally studied in \cite{Rip04}. Since we employ a one-dimensional FDTD scheme, the strength of the current $I$ required to produce a magnetisation reversal in our model is lower than that in \cite{Rip04}.

\subsection{Results}
In Fig.~\ref{Fig1}b we plot the dynamics of the $M_z$ component of the magnetisation vector ${\bf M}$ in the ``free'' layer as a function of the electric current strength for different value of the applied magnetic field $H_0$. The trajectories of ${\bf M}$ for $H_0=8$\,kOe computed at the simulated time of 1\,ns are shown in Fig.~\ref{Fig4}b--c. Figure~\ref{Fig1}b demonstrates that the magnetisation dynamics correctly approximates the sigmoid probabilistic PR curve in Fig.~\ref{Fig1}a but the strength of the driving electric current plays the role of parameter $A$.

The experimental curves in Fig.~\ref{Fig3} can be regarded as the particular cases of the sigmoid function and they can be simulated using the magnetisation reversal observed at different values of $H_0$ (for example, compare the experimental CDF curve in Fig.~\ref{Fig3}c.ii with the magnetisation reversal curve for 9\,kOe in Fig.~\ref{Fig1}b). We also note that it is computationally straightforward to use our model to approximate curves whose shape deviates from a sigmoid function. To demonstrate this possibility, in Study~2 we unleash the potential of the magnetisation reversal model applying it to the analysis of a more complex experimental dataset.

\section{Study~2}
We further validate our model using a dataset obtained from a ``Deal or No Deal'' video game based on a popular TV show, where contestants choose boxes with concealed cash amounts and negotiate with a banker to accept an offer or keep opening boxes. The aim is to secure the highest prize, avoiding smaller sums. Once a box is opened, its prize is shown and removed from potential winnings. Essentially, players choose between risky lotteries and certain cash offers. They also may exchange their box for any unopened one in the game. The decision to swap the box or to stick to the original one enables us to study whether the participants behave according to EUT or to Cumulative Prospect Theory (CPT). A previous study \cite{blavatskyy2010endowment} that used TV show data from three countries demonstrated that: (i)~CPT players are more likely to stick to their original choice of the box due to the embedded assumption of loss aversion and (ii)~EUT players are indifferent between swapping and sticking.

\begin{figure}
 \includegraphics[width=0.9\textwidth]{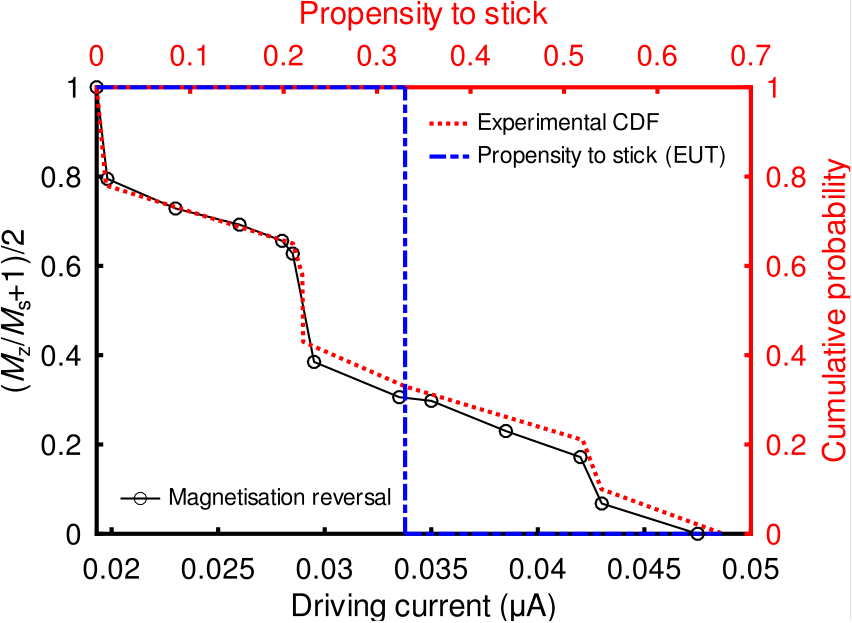}
 \caption{Top $x$ and right $y$ axes:~Cumulative probability distribution function for sticking decisions in the video game. The dashed straight lines denote the EUT propensity to stick. The dotted curve corresponds to the experimental data. Bottom $x$ and left $y$ axes:~The circular markers denote the simulated magnetisation reversal. The thin solid line is the guide to the eye only.\label{Fig5}}
\end{figure}

We recruited 78 study participants (members of the general public over 18 years of age, informed consent was obtained from all participants as per the procedure used in Study~1) who made a total of 1,698 game decisions, including 486 swap or stick decisions. The fact that 63\% of decisions were stick decisions and 37\% were swap decisions demonstrated a higher consistency with EUT than with CPT. However, Cumulative Distribution Function (CDF) of swap or stick decisions revealed an intriguing result: while EUT (the straight dashed lines in Fig.~\ref{Fig5}) predicts an approximately equal split between swapping or sticking decisions, the distribution of swap or stick decisions (the dotted curve in Fig.~\ref{Fig5}) cannot be described by EUT.

Since the magnetisation dynamics depends on the static field $H_0$ (Fig.~\ref{Fig1}b), in Fig.~\ref{Fig5} we plot a magnetisation reversal that is recast as $(M_z/M_s+1)/2$ for the sake of comparison with the experimental data. The resulting curve satisfactorily reproduces the experimental decision-making behaviour demonstrated by the players of the game. The values of $H_0$ corresponding to the data points of this curve are, from left to right: 15, 6, 8.75, 9.6, 9.95, 10, 10, 10.75, 11, 11.52, 11.99, 12.1 and 12.5\,kOe.

\section{Discussion}
The functions of a human brain underpinning the phenomena like PR can be comprehended in detail through neuroscience methods \cite{Sha21}. Nonetheless, understanding human behaviour is pivotal not just for neuroscience but also for fields of finance, politics, religion and sociology. Thus, there is a demand for phenomenological models that can forecast decisions at both individual and societal levels quickly and efficiently.

These models should not necessarily be based on neurobiological insights, mirroring the human cognition (drawing parallels with biological brain processes might even be unhelpful \cite{Tad16}). The primary criteria for evaluating a model of decision-making should be the ability of the model to accurately explain and predict imprecise and noisy decisions. The model we introduce in this paper follows this philosophy, aligning with the ongoing work in decision theory where models are not necessarily capturing human brain functions \cite{wakker2010prospect}.

Importantly, our model is not driven by experimental datasets. In fact, the psychological data presented in this paper are needed only to validate the model. The validation process requires relatively compact datasets and its main aim is to establish a range of model parameters that can be used in further studies. Once suitable parameters have been established, the model can produce new results independently of experimental data.

This approach has certain limitations, including the need to adjust the model parameters to analyse a previously unconsidered social or economical scenario (a more comprehensive discussion of the perceived limitations of the proposed model is given in Sect.~\ref{FutRes}). However, the adjustment process should be relatively simple since it requires one to validate the model against another compact set of experimental data. Overall, results produced by a properly adjusted model can help reduce the need to conduct complex experiments involving many human participants.

STNO has been a foundational component in neuromorphic computing hardware systems that emulate the functions of the human brain \cite{Fur18, Rio19}. The operational principles of STNO obey the laws of quantum mechanics \cite{Bra22}, even though our model's Eq.~(\ref{eq:eq1}) is classical. Quantum-mechanical models have demonstrated superior capacity in depicting decision-making compared with classical models \cite{Bus12}. Therefore, leveraging a realistic STNO structure in our model hints at the potential of STNOs to act as a neuromorphic system for simulating human decision-making processes. The so-designed system can be integrated with a reservoir computing algorithm \cite{Mak23_review}, resulting in a machine learning technique that has an enhanced learning efficacy compared with classical systems \cite{Dud23} and also can simulate the human decision-making. 

Other quantum-mechanical systems can replicate decision-making \cite{Nar14}. However, the magnetisation reversal in STNO naturally extends important models of opinion shifts in social networks based on spin direction reversal \cite{Red19}. Even though those models adopted the spin concept from physics, the processes causing the spin changes were largely synthetic. Thus, integrating a genuine model of magnetisation reversal into the opinion shift model \cite{Red19} will enhance the efficacy of the latter, enabling it to analyse diverse decision-making scenarios.

\section{Future Research\label{FutRes}}
The current work is a plausible attempt to phenomenologically describe psychological and decision-making phenomena using a purely physical model. Although this approach is mathematically similar to models belonging to the quantum decision theory group, our model was developed independently, exploiting the ability of classical physics to successfully explain certain quantum-mechanical phenomena.

Admittedly, our model cannot capture the complexity of cognitive processes in full. However, creating such a model goes well beyond the scope of this paper. Yet, as famously written by Box and Draper, the approximate nature of the model must always be borne in mind \cite{Box87}. In other words, any model has its advantages and disadvantages and its true performance can be demonstrated only in tests involving real-life data. Supporting this point of view and recognising the benefits of a comparison of different models \cite{Bus00}, in this paper we establish the ground truth by benchmarking our model against large datasets obtained in diverse experimental scenarios, which a standard research approach in the fields of physics and engineering. It is also likely that additional limitations of our model will be revealed at later stages, when the model will be used by different users. This process is natural and it will help us and other researchers following our work improve the model.  

Importantly, since we aimed to make our model accessible to a broad range of non-experts in physics, we deliberately kept the model equations as simple as possible. Hence, future research could explore more intricate physical processes and more advanced quantum-physical models that may offer a closer analogue to the multifaceted nature of human cognition and decision-making. Indeed, the experimental setups, such as those employed in our ``Deal or No Deal'' study, provide valuable insights but also come with limitations regarding, for example, the range of decision-making scenarios they can simulate. Further studies could employ more diverse and real-life decision-making contexts, enhancing the applicability and robustness of our findings.

This study predominantly focuses on an external, observable behaviour perspective. Integrating other kind of insights, especially from fields like macroeconomics, could provide a more holistic understanding of decision-making processes, bridging the gap between physical models and the neurological underpinnings of human behaviour. The inclusion of these factors in our model should not represent significant technical difficulties. Indeed, strictly speaking, if a decision-making process can be represented by a continuous function at least in principle then our model or its modifications should be able to approximate it with a reasonable accuracy. Yet, our physical model can naturally incorporate noise, interference and dissipation, all defined physically, which are the processes that were previously explored in models of cognition and decision-making \cite{Khr06, Aer22, Aer22_1}.  

The computational models employed in psychology, while effective, may not fully capture the dynamism and adaptability of human decision-making. Future research could incorporate machine learning and AI to develop more sophisticated models that can dynamically adapt and learn from new data, mirroring the evolving nature of human preferences. One example of a relevant study is our recent work on a quantum-inspired neural network model of optical illusions that exploits a chaotic physical system as a network \cite{Mak24}. The cited work aligns with a broader collective effort to create neural networks that could predict complex socio-economic processes using the laws of physics (for a review of methodological aspects see, e.g.,~\cite{Mak23_review}). 

Our study primarily considers decision-making from a psychological and physical perspective, potentially overlooking the influence of socio-cultural factors, which may be of interest. Subsequent research could integrate these variables, examining how cultural, social, and environmental factors interplay with the cognitive processes in decision-making. We are currently working in this direction and the results of that project will be reported elsewhere. 

By addressing these limitations, future research can not only refine and expand upon our current findings but also contribute to a more nuanced and comprehensive understanding of human decision-making processes. More specifically, the current research can lead to a deeper integration of physical sciences with behavioural economics and psychology. Future research could contribute by developing comprehensive models that further incorporate principles from quantum mechanics, statistical physics and complexity science to capture the nuanced aspects of human decision-making, as exemplified by neural network models empowered by quantum technologies \cite{Mak24}. There is also an opportunity to explore the neurobiological underpinnings of the decision-making phenomena modelled by physical systems. Indeed, future studies could use neuroimaging techniques to validate the predictions of physical models against observed brain activity during decision-making tasks (see, e.g., the works on artificial neural network that mimic the operation of a biological brain \cite{Rao22} and relevant works on physics-informed machine learning models \cite{Nak22, Mak23_review}).

Finally, advanced computational techniques like machine learning and AI could be employed to refine the models based on large datasets. In this context, future research could focus on the development of algorithms that learn from behavioural data to predict decision-making patterns and preference changes over time. A number of papers developing relevant ideas have already been published \cite{Bus17, Khr18, Khr20, Sal23}. Future research could utilise such models to simulate the effects of policy changes on public behaviour: by understanding how individuals might react to different policy interventions more effective and targeted strategies can be developed. Indeed, there have been several demonstrations of the possible application of quantum-mechanics in political studies, especially in the realm of international relations studies \cite{Jak15, All18}. Importantly, the cited works present rather a general philosophical discussion of such models, thereby highlighting unexplored fields of research where physical model of decision-making can find specific practical applications.

\section{Conclusions}
We demonstrated that the psychological phenomenon of PR can be modelled using the physical phenomenon of magnetisation reversal. Using a spin transfer torque as the mechanism of magnetisation reversal, we revealed the ability of the magnetisation reversal model to capture complex individual's decisions made in psychological experiments. We have also shown that our model extends beyond PR and encompasses a wide range of decision-making regularities, reaffirming the observation that human preferences are imprecise.

\subsection {Implications on Literature}
This research contributes to a dynamic and ever-evolving body of literature that challenges traditional economic and psychological theories. It encourages the development of new theories that accommodate the stochastic and probabilistic nature of human decision-making. The use of physical models to explain psychological phenomena can inspire new research methodologies that are more robust and predictive of real-world behaviours. The findings demonstrate the value of cross-disciplinary research and can stimulate collaborative studies that bridge gaps between fields such as cognitive psychology, economics, physics and computer science.

\subsection {Implications on Practice and Policy}
In practice, these models can inform the design of behavioural interventions that are more aligned with how people actually make decisions, potentially increasing their efficacy. The ability to more accurately predict human behaviour can revolutionise economic forecasting and financial planning, providing tools for better risk assessment and market analysis. Policymakers can use insights from this research to craft policies that consider the probabilistic nature of human decisions, leading to more nuanced and effective public policies. The interdisciplinary nature of this research can also influence educational curricula, promoting a more integrative approach to teaching economics, psychology and physics.

While our work bolsters the exploration of physics as a lens to comprehend human psychology, its tangible implications are profound: the model holds promise for designing neuromorphic hardware systems rooted in nanomagnetic devices. Such interdisciplinary convergence has the potential to reshape the way we interface with both the human mind and advanced computational systems.

The implication of magnetisation reversal as a model for human decision-making not only accentuates the universality of certain principles across ostensibly disparate domains but also underscores the rich tapestry of relationships waiting to be unearthed between the physical and cognitive realms. For example, the integration of these findings can stimulate a paradigm shift in how we design and conceptualise next-generation cognitive devices. Moreover, the ability of physical effects to emulate cognitive processes suggests a future where the boundaries between cognition and artificial intelligence blur, leading to new possibilities for collaborative intelligence.

\subsection {Implications on Interdisciplinary Research in Physics}
Last but not least, the present work contributes to a more than four decades long endeavour to establish the field of sociophysics \cite{Gal22}. Although sociophysics has maturated as an independent discipline, it has been unable to find a secure place in specialised physical journals and university classrooms yet. The present work demonstrates that a physical model developed to investigate the fundamental physical effect of magnetisation reversal can be applied in the domain of social sciences without the need to modify the model parameters. While some physicists may find this result to be contradictory, our findings demonstrate that the actual frontiers of physics are located well beyond our current understanding and that the point where a physical topic becomes a non-physical one is more difficult to define than is immediately apparent.

\section*{Ethics statement}
This work was approved by Humanities and Social Sciences Research Ethics Committee (HSSREC) of the University of Warwick (approval number~31/13-14). Upon arrival to the laboratory, each participant was provided with the study information sheet and a consent form. Informed consent was obtained from all participants in the study. All participants were adults (18 years of age or older). No minors participated in the experiments; therefore, obtaining parental/guardian consent was not applicable to this work.

\section*{Data Availability}
All data relevant to this work are presented in this paper. Large-size relevant raw data are uploaded to GitHub: \url{https://github.com/BehaviouralDataScience/Magnetisation-Reversal-Model} alongside the software used to process them.

\section*{CRediT authorship contribution statement}
Ivan S.~Maksymov: Writing--original draft, Writing--review \& editing, Visualization, Methodology, Investigation, Formal analysis, Conceptualization. 

Ganna Pogrebna: Writing--original draft, Writing--review \& editing, Visualization, Project administration, Methodology, Investigation, Formal analysis, Conceptualization.\\

\noindent All authors gave final approval for publication and agreed to be held accountable for the work performed therein.

\section*{Competing interests}
The authors declare that they have no known competing financial interests or personal relationships that could have appeared to influence the work reported in this paper.

\section*{Declaration of AI use}
We have not used AI-assisted technologies in creating this article.

\section*{Funding}
We received no funding for this study.

\section*{Acknowledgements}
Not applicable.

\bibliographystyle{elsarticle-num}
\bibliography{refs}

\end{document}